\documentclass[10pt,conference]{IEEEtran}

\usepackage{mystyle}

\hypersetup{
	pdftitle={Assemble Foundation Models for Automatic Code Summarization},
	pdfauthor={Jian Gu, Pasquale Salza, Harald C. Gall},
	pdfkeywords={transfer learning, adaptive scheme, Transformer, Gaussian noise, code summarization}
}

\DeclareAcronym{api}{
	short = API,
	long = {Application Program Interface}
}

\DeclareAcronym{awgn}{
	short = AWGN,
	long = {additive white Gaussian noise}
}

\DeclareAcronym{vae}{
	short = VAE,
	long = {Variational AutoEncoder}
}

\DeclareAcronym{bert}{
	short = BERT,
	long = {Bidirectional Encoder Representations from Transformers}
}

\DeclareAcronym{roberta}{
	short = RoBERTa,
	long = {Robustly optimized BERT approach}
}

\DeclareAcronym{ast}{
	short = AST,
	long = {Abstract Syntax Tree}
}

\DeclareAcronym{bpe}{
	short = BPE,
	long = {Byte-Pair Encoding}
}

\DeclareAcronym{cfg}{
	short = CFG,
	long = {Control Flow Graph}
}

\DeclareAcronym{dcg}{
	short = DCG,
	long = {Discounted Cumulative Gain}
}

\DeclareAcronym{gpt}{
	short = GPT,
	long = {Generative Pretrained Transformer}
}

\DeclareAcronym{ir}{
	short = IR,
	long = {Information Retrieval}
}

\DeclareAcronym{lstm}{
	short = LSTM,
	long = {Long Short-Term Memory}
}

\DeclareAcronym{clm}{
	short = CLM,
	long = {Casual Language Modeling}
}

\DeclareAcronym{mlm}{
	short = MLM,
	long = {Masked Language Modeling}
}

\DeclareAcronym{mem}{
	short = MEM,
	long = {Multimodal Embedding Model}
}

\DeclareAcronym{cp}{
	short = CP,
	long = {Continuous Pretraining}
}

\DeclareAcronym{if}{
	short = IF,
	long = {Intermediate Finetuning}
}

\DeclareAcronym{mmpf}{
	short = MMPF,
	long = {Massive Multitask Pre-Finetuning}
}

\DeclareAcronym{aif}{
	short = AIF,
	long = {Adaptive Intermediate Finetuning}
}

\DeclareAcronym{mrr}{
	short = MRR,
	long = {Mean Reciprocal Rank}
}

\DeclareAcronym{ndcg}{
	short = NDCG,
	long = {Normalized Discounted Cumulative Gain}
}

\DeclareAcronym{nlp}{
	short = NLP,
	long = {Natural Language Processing}
}

\DeclareAcronym{nlp_pt}{
	short = NLP\textsubscript{PT},
	long = {Next Line Prediction}
}

\DeclareAcronym{nmt}{
	short = NMT,
	long = {Neural Machine Translation}
}

\DeclareAcronym{nsp}{
	short = NSP,
	long = {Next Sentence Prediction}
}

\DeclareAcronym{rnn}{
	short = RNN,
	long = {Recurrent Neural Network}
}

\DeclareAcronym{cnn}{
	short = CNN,
	long = {Convolutional Neural Network}
}

\DeclareAcronym{tf-idf}{
	short = tf-idf,
	long = {term frequency–-inverse document frequency}
}

\DeclareAcronym{anova}{
	short = ANOVA,
	long = {ANalysis Of VAriance}
}

\DeclareAcronym{da}{
	short = DA,
	long = {Domain-Adaptive}
}

\DeclareAcronym{ta}{
	short = TA,
	long = {Task-Adaptive}
}

\DeclareAcronym{ma}{
	short = MA,
	long = {Multiphase Adaptive}
}

\DeclareAcronym{ca}{
	short = CA,
	long = {Concept Annotation}
}

\DeclareAcronym{ce}{
	short = CE,
	long = {Concept Extrapolation}
}

\DeclareAcronym{ci}{
	short = CI,
	long = {Concept Interpolation}
}

\DeclareAcronym{gru}{
	short = GRU,
	long = {Gated Recurrent Unit}
}

\DeclareAcronym{sota}{
	short = SOTA,
	long = {state-of-the-art}
}

\DeclareAcronym{lcs}{
	short = LCS,
	long = {Longest Common Sequences}
}

\begin{document}

\bstctlcite{IEEEexample:BSTcontrol}

\title{Assemble Foundation Models\\for Automatic Code~Summarization}
\author{

\IEEEauthorblockN{Jian Gu}
\IEEEauthorblockA{%
University of Zurich\\
Zurich, Switzerland\\
e-mail: \href{mailto:gu@ifi.uzh.ch}{gu@ifi.uzh.ch}%
}

\and

\IEEEauthorblockN{Pasquale Salza}
\IEEEauthorblockA{%
University of Zurich\\
Zurich, Switzerland\\
e-mail: \href{mailto:salza@ifi.uzh.ch}{salza@ifi.uzh.ch}%
}

\and

\IEEEauthorblockN{Harald C. Gall}
\IEEEauthorblockA{%
University of Zurich\\
Zurich, Switzerland\\
e-mail: \href{mailto:gall@ifi.uzh.ch}{gall@ifi.uzh.ch}
}

}

\maketitle
\begin{abstract}
Automatic code summarization is beneficial to daily software development since it could help reduce the requirement of manual writing. Currently, artificial intelligence is undergoing a paradigm shift. The foundation models pretrained on massive data and finetuned to downstream tasks surpass specially customized models. This trend inspired us to consider reusing foundation models instead of learning from scratch. Thereby, we propose a flexible and robust approach for automatic code summarization, based on neural models. We assemble available foundation models, such as CodeBERT and GPT-2, into a single neural model named AdaMo. Moreover, we utilize Gaussian noise as the simulation of contextual information to optimize the latent representation. Furthermore, we introduce two adaptive schemes from the perspective of knowledge transfer, namely continuous pretraining and intermediate finetuning, and design intermediate stage tasks for general sequence-to-sequence learning. Finally, we evaluate AdaMo against a benchmark dataset for code summarization, by comparing it with state-of-the-art models.
\end{abstract}


\section{Introduction}
\label{sec:introduction}

In the process of software development and maintenance, decent comments are crucial to program comprehension and could reduce the burden of directly interpreting the code~\cite{le_deep_2020}.
By reading and writing code comments, software developers instantly load and save the working context, such as requirements and implementation details.
However, it is laborious to manually write comments in a consistent and decent style and keep them synchronized with the code simultaneously.
It seems inevitable to study the automatic way of generating high-quality code comments, namely \enquote{code summarization}, which could save massive effort and time.

\emph{Code summarization} is the task of generating natural language descriptions for the given code snippets.
In practice, it serves numerous daily activities in software development and maintenance, such as recording implementation details, templatizing package descriptions, and describing code changes for version updates~\cite{zhu_automatic_2019}.
Its automatic solutions could improve productivity in software activities effectively.
There are several solutions proposed for automatic code summarization, and among them, neural models are the most valued for their unique generative ability.
These machine learning solutions significantly rely on the knowledge learned from a large code corpus.
Correspondingly, neural models require massive data and intensive computational resources.

In this paper, we build a transfer learning model \textbf{\adamo} composed of relevant foundation models for code summarization and also introduce Gaussian noise to enhance the latent representations.
Our experimental results showed that \adamo defeated \ac{sota} models.

Recent work tend to introduce structure information extracted from the parsed results of code data, \eg, \ac{ast} and \ac{cfg}, as a modality complement.
The representatives are \ac{sota} models, namely \basts~\cite{lin_improving_2021} and \sit~\cite{wu_code_2021}.
Instead of lightly processing the code data as a token sequence, they parse it into tree or graph forms, thereby taking heavy computation consumption.
Compared with \ac{sota} models, our approach is merely based on the token sequence.
As token-based neural models proved to be effective in handling sequential data, especially in \ac{nlp}, we want to leverage on them and see how far we can go.
This brings us an advantage over baseline models that probably requires fully parsable code, whereas our model works as well on corrupted or partial code.

To make up for the disadvantages that our model only learns from less information, our model emphasizes reusing pretrained checkpoints for better parameter initialization~\cite{han_pretrained_2021}. With the popularity of transfer learning, foundation models are also emerging~\cite{bommasani_opportunities_2021}.
They are generally trained on intensive data at scale and then adapted to various target tasks.
Based on that, we leverage pretrained models instead of training a model from scratch.
We assemble the representative encoder model and decoder model, \eg, \bert~\cite{devlin_bert_2019} and \gpt~\cite{radford_improving_2018}, and then train their assembly.
Furthermore, considering that adaptive schemes are already proposed in transfer learning, as the complement to the standard \enquote{\emph{pretraining} and then \emph{finetuning}} paradigm, we further design intermediate tasks to adopt continuous pretraining and intermediate finetuning on code summarization.

\smallskip
\noindent
To summarize, the main contributions of this paper are:
\begin{itemize}
    \item proposing an effective and straightforward approach for code summarization, by assembling foundation models;
    \item adopting the Gaussian noise emitter as a simulation of contextual information for better latent representations;
    \item introducing adaptive transfer learning schemes as options to further raise the upper bound of model performance;
    \item designing intuitive intermediate stage tasks for code summarization and general sequence-to-sequence learning.
\end{itemize}

\smallskip
\noindent
The rest of this paper is structured as follows.
In \cref{sec:background}, we introduce the background knowledge.
In \cref{sec:approach}, we present our transfer learning model and adaptive schemes.
\Cref{sec:experiments} describes the experimental setup, whereas \cref{sec:results} presents and discusses the results.
\Cref{sec:related_work} surveys the related work, and \cref{sec:conclusions} concludes with a summary of the findings and contributions and an outlook on future work.

\section{Background}
\label{sec:background}

In this section, we describe the background of our work, such as \transformer models, \eg, \bert~\cite{devlin_bert_2019} and \gpt~\cite{radford_improving_2018}, transfer learning and adaptive schemes.

\subsection{\transformer Models}
\label{sec:background:transformer}

The standard \transformer is an encoder-decoder neural model that merely relies on the attention mechanism~\cite{vaswani_attention_2017} as the main component.
We call other models inspired by this simple and effective way of building networks as \transformers.

The architecture of \transformer is a stack of six encoder layers mingled with a stack of six decoder layers.
In the encoder layer, there are a self-attention sublayer and a feed-forward network.
Instead, in the decoder layer, there is an extra encoder-decoder attention sublayer.
The self-attention sublayer is for the connections within encoder layers or decoder layers, while the encoder-decoder attention sublayer is for the connections between the last encoder layer and each decoder layer.
Both the self-attention sublayer and encoder-decoder attention sublayer adopt multiple attention units to quantify the importance of each part of the input data in the same manner.

Due to the superiority of \transformers in the computation complexity and flexibility over prior models, they are popular, especially in the field of \ac{nlp}.
Besides, its capability of computation parallelization allows more intense use of massive data, which led to the development of large-scale pretrained models, namely foundation models~\cite{bommasani_opportunities_2021}.
In this paper, we consider two examples of \transformers in particular: \bert~\cite{devlin_bert_2019} for discriminative tasks, and \gpt~\cite{radford_improving_2018} for generative tasks.

\paragraph{\acf{bert}}
\bert~\cite{devlin_bert_2019} is the typical encoder model inspired by \transformer and follows the idea of merely using the attention mechanism to build the neural model.

In terms of model architecture, \bert stacks \num{12} encoder layers (\num{24} layers for its large version), but no decoder layers.
\bert-like models are used for \emph{discriminative tasks}.
Each encoder layer of \bert consists of a self-attention sublayer and a feed-forward network, the same as \transformer.
In the self-attention sublayer, each token can attend context to its left and right, so the attention is referred to as \enquote{bidirectional}.

\ac{mlm} is a common objective to train \bert-like models.
In \ac{mlm}, the model is trained to predict \num{15}$\%$ randomly masked tokens based on the entire context, namely the tokens that occurred on both two sides.
The \ac{mlm} objective works in pair with the bidirectional self-attention, to guide the model to learn a decent representation of the input.
RoBERTa~\cite{liu_roberta_2019}, as its optimized version, achieves further improvements by dynamically masking tokens and offering a bigger data capacity. \codebert~\cite{feng_codebert_2020} is a specialized \roberta for the representation of code data.

\paragraph{\acf{gpt}}
\gpt~\cite{radford_improving_2018} is the typical decoder model inspired by \transformer that mainly uses the attention mechanism to build the neural model.

In terms of model architecture, \gpt only stacks \num{12} decoder layers (\num{24}, \num{36}, or \num{48} layers for its medium, large, and extra large versions), but no encoder layers.
\gpt-like models are used for \emph{generative tasks}
Each decoder layer in \gpt consists of a masked self-attention sublayer and a feed-forward network, which is different from \transformer.
Besides, there is no encoder-decoder attention sublayer.
In the masked self-attention sublayer, each token only attends context to its left.
Thus the attention is referred to as \enquote{constrained}.

The only self-supervised training objective of \gpt is \ac{clm}.
In \ac{clm}, the model learns to predict \num{15}$\%$ randomly masked tokens when given the partial context, namely the tokens that occurred only on their left sides.
The \ac{clm} objective works in pair with the constrained self-attention, to guide the model to generate a proper sentence as the output.
Its enhanced versions, \ie, \gpttwo~\cite{radford_language_2019}, show stronger capability in text generation.

\subsection{Transfer Learning and Adaptive Schemes}
\label{sec:background:adaptive}

Different from general learning modes, \emph{Transfer Learning} is the learning paradigm of transferring the knowledge learned from other data to the new data for given domains or tasks~\cite{pan_survey_2010}.
It relaxes the requirements on data amount, data distribution, and computation capability~\cite{niu_decade_2020}.
Besides the conventional \enquote{\emph{pretraining} and then \emph{finetuning}} practice, adaptive schemes are applicable to transfer learning models.

The intuition behind transfer learning is that the beforehand self-supervised training for data representation promises general and reliable initialization, \ie, knowledge of data distribution, and then benefits the performance of specific downstream tasks, such as raising the upper bound or accelerating convergence.
Compared with starting from scratch, transfer learning could improve the sample efficiency since it reduces the consumption required on data and computational resources~\cite{georgekarimpanal_selforganizing_2018}.

Based on transfer learning, models for various tasks could origin from the same pretrained language model.
As an extended definition of pretrained language model, the model that is \emph{\enquote{trained on broad data at scale and can be adapted to a wide range of downstream tasks}} is named \enquote{foundation model}~\cite{bommasani_opportunities_2021}.
In \ac{nlp}, representative examples include \bert and \gpt, which are trained on large corpora of text and then adapted to a wide range of downstream tasks, \eg, machine translation, question answering, and sentiment analysis.

The standard methodology of transfer learning is \emph{pretraining} a model on a large corpus of unlabeled data and then \emph{finetuning} it on a small supervised dataset.
In the pretraining stage, the model is usually trained in a \emph{self-supervised learning} manner, where the unlabeled data is sufficient for the objective, therefore the pretraining data is usually extensive and readily available.
Instead, during the finetuning stage, the model is trained in a \emph{supervised learning} manner where the ground truth is required.
The data used for the target (downstream) task is supervised data, and its quality matters the most, so the data amount is usually limited and the cost is more expensive.
Even though it has been conventional to pretrain and then finetune models, there are still strategies proposed to further improve the data adaptability to target domains or tasks, which mainly work between the usual pretraining and finetuning stages.

\emph{\acf{cp}} is defined as tailoring a model to another data domain, or even the data of a target task, via the second phase of pretraining~\cite{gururangan_don_2020}.
It has been shown that domain-adaptive pretraining, namely adapting the model to the data of the same domains, improves the performance.
Similarly, adapting the model for designated tasks, called task-adaptive pretraining, leads to performance gain as well.
Moreover, multiphase adaptive pretraining, \eg, domain-adaptive training followed by the task-adaptive one, promises a larger gain.

\emph{\acf{if}} benefits a pretrained model by introducing intermediate tasks during the additional training stage, as the warmup activities before training for the target task~\cite{pruksachatkun_intermediatetask_2020}.
However, the characteristics of intermediate tasks could affect the effectiveness of adaptive finetuning.
Divided by difficulty, intermediate tasks could be either simple or complex.
The simple intermediate tasks are close to learning the low-level skills such as preserving the raw content and detecting the shallow attributes, such as verb tenses or sentences length.
They would only affect the model performance slightly.
In contrast, complex intermediate tasks are generally rather beneficial to promote the model, \eg, natural language inference~\cite{storks_recent_2019}, and question answering~\cite{min_knowledge_2019}.
Thus, they expect the model to have strong capabilities such as perceiving interrelations.

\section{Approach}
\label{sec:approach}

This section describes the model architecture and its components and how we applied transfer learning.
Then, we introduce adaptive schemes~\cite{ruder_recent_2021}, \eg, continuous pretraining, and intermediate finetuning.

\subsection{Model Architecture}
\label{subsec:approach:model}

Our transfer learning model follows the encoder-decoder architecture for general sequence-to-sequence learning~\cite{sutskever_sequence_2014}, which has stacked encoding layers as the encoder and stacked decoding layers as the decoder.

Instead of training from scratch, building a model by adopting existing foundation models supports reusing the pretrained weights for model initialization~\cite{rothe_leveraging_2020}.
Thereby, we propose such a model that adopts \codebert~\cite{feng_codebert_2020}, \eg, a specialized \roberta for code representation, as the encoder, and \gpttwo~\cite{radford_language_2019}, \eg, an enhanced \gpt for text generation, as the decoder.
In addition, we place an \ac{awgn} emitter between the encoder and the decoder as an option to further optimize the latent representation.
We name such a transfer learning model \textbf{\adamo}, standing for \ul{Ada}ptive \ul{Mo}del with reference to the adaptive schemes it supports.

\begin{figure}[!tb]
    \centering
    \includegraphics[width=0.9\linewidth]{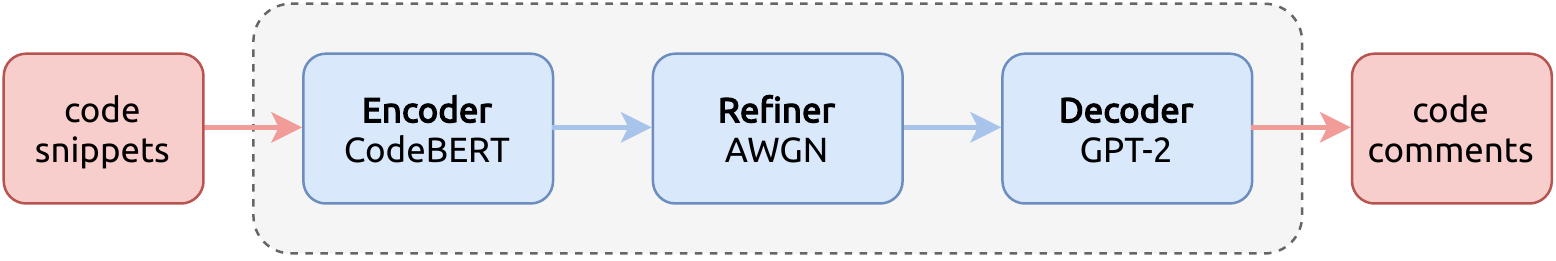}
    \caption{The \adamo architecture.}
    \label{fig:model}
\end{figure}

Considering the input data is code snippets, and output data is code comments, our model adopts \codebert as the encoder and \gpttwo as the decoder.
\codebert can capture the general representations of code data, and
similarly, \gpttwo performs well in English text generation.
We make no changes to the metadata of the encoder model and decoder, such as model structure and hyperparameters.
Besides, the noise emitter between encoder and decoder is expected to generate Gaussian noise as the rough simulation of contextual information of the input data~\cite{li_does_2020}.
As shown in \cref{fig:model}, the red blocks are input data and output data, and the dashed block is the architecture of our model, whose components are represented as blue blocks.

We define the processor manipulating the latent representations between encoder and decoder as \emph{Refiner}.
Furthermore, we define the latent representations of encoder and decoder as \emph{Parallel Representations}.
In our model architecture, encoder and decoder both reuse the network design of existing language models and are initialized with the pretrained weights.
Therefore the parallel representations of both sides are not naturally coordinated with each other.
The refiner is thereby introduced to optimize the latent representations and reduce their incoordination.

Prior work in natural language translation found it could be equivalent to introducing multiple encoders to capture contextual information when accumulating Gaussian noise to the latent representations~\cite{li_does_2020}.
Gaussian noise is a basic noise model, following the normal distribution, commonly used to mimic the effect of random processes in nature.
Considering both code and text data are sequential, and the context information matters as well in code data, we adopt the \ac{awgn} emitter to simulate the context-aware setting of multiple encoders.

\begin{figure}[!tb]
    \centering
    \includegraphics[width=0.9\linewidth]{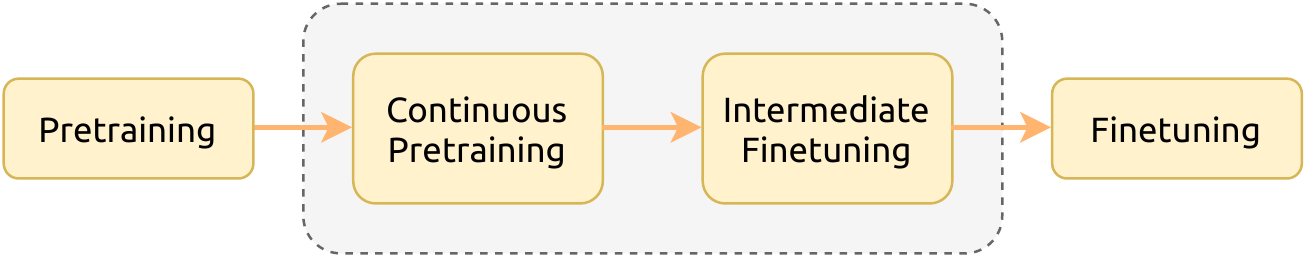}
    \caption{The Adaptive schemes.}
    \label{fig:scheme}
\end{figure}

Our \adamo model is mainly composed of a \transformer encoder and decoder, and its architecture naturally supports practices applicable in transfer learning, such as the adaptive schemes.
As shown in \cref{fig:scheme}, we introduce adaptive schemas, \eg, continuous pretraining and intermediate finetuning, between the common pretraining and finetuning stages, to make the model more adaptive to the data of specific domains or tasks.
The adaptive schemes are expected to bring improvements by raising the upper bound of model performance.
They do not necessarily appear simultaneously in our approach, but their combination is feasible.

\subsection{Adaptive Schemes}
\label{subsec:as}

Based on the standard transfer learning paradigm of \enquote{\emph{pretraining} and then \emph{finetuning}}, we introduce adaptive schemes for better data adaptability to target domains or tasks, \eg, continuous pretraining and intermediate finetuning.

For continuous pretraining, irrespective whether it is \ac{da} or \ac{ta}, there exist no obvious differences with the general pretraining stage.
However, \ac{da} pretraining specifies that the unlabeled data to be used must come from a related domain, while \ac{ta} pretraining directly utilizes the unlabeled data of the given task.
For instance, assuming that we would use a dataset of Physics papers to finetune a pretrained language model, depending on the concrete data to use, continuous pretraining is identified as domain-adaptive or task-adaptive.
If the unlabeled data is not from the given dataset but another related dataset, such as the dataset of general scientific papers, it is \ac{da} pretraining.
On the contrary, if the unlabeled data is from the given dataset, it is referred to as \ac{ta} pretraining.

For both \ac{da} and \ac{ta}, we set \acf{mlm} as the objective of the encoder, and \acf{clm} as the objective of the decoder.
Both \ac{mlm} and \ac{clm} involve randomly masking parts of the sequential data, then training the model to predict the missing tokens correctly.
For the prediction, \ac{mlm} allows the model to consider the entire context, \ie, the data occurring to both left and right sides.
Instead, \ac{clm} only allows the model to consider the partial context, \ie, the data occurring to the left side only.
Specifically, for \ac{da} pretraining, we pretrain the model with massive unlabeled data from another related dataset, separately train only either encoder or decoder, or train them both.
For \ac{ta} pretraining, we pretrain the model almost in the same way but directly using the unlabeled data from the given dataset.

\begin{custombox}
    \textbf{\Acl{da}:} The adaptation to the distribution of the related domain of the target data in the model migration.
    
    \smallskip\noindent
    \textbf{\Acl{ta}:} The adaptation to the distribution of the target data when the model migrates from the source data.
\end{custombox}

To the best of our knowledge, there exist no ready-made intermediate tasks designed for code summarization yet, and none for general sequence-to-sequence learning.
Inspired by text editing~\cite{mallinson_felix_2020}, we propose that the summarization task could be seen as the fusion of two novel stage tasks: one is to insert tokens that occurred in the target text but not in the source text; the other one is to delete tokens that occurred in the source text but not in the target text.
Also, the stage tasks concern the order of tokens as well because reordering tokens is a mandatory implicit operation.
Considering this, we regard the changes in both the occurrence and order of tokens in the summarization task as concept shifts.
We name the first stage task as \acf{ce} and the second one as \acf{ci}.
In addition, one intuitive but straightforward idea is to directly annotate the tokens in the target text with different marks based on whether they are available in the source text.
We propose this idea as the comparison task and name it as \acf{ca}.
Not limited to \emph{code} summarization, \ac{ca}, \ac{ce}, and \ac{ci} are applicable to \emph{text} summarization, even general sequence-to-sequence learning.

For a given sentence, both \ac{ce} and \ac{ci} cover the operation of reordering tokens, however, \ac{ce} is more like a typical complex task while \ac{ci} seems relatively simple.
The reason is that the model has to associate and append tokens that never appeared in the sentence in the case of \ac{ce}.
Instead, for \ac{ci} the model merely determines whether each existing token in the sentence should be kept or dropped.
\acf{ca} is a bit different from others because it only annotates tokens and introduces no information about the order of the target text.
We regard it as the most straightforward intermediate task.
The number of intermediate tasks is suggested to be \num{1} because when the amount of involved tasks is less than a critical point, \eg, \num{15}, then the fewer tasks, the better the performance~\cite{aghajanyan_muppet_2021}.

\begin{custombox}
    \textbf{\Acl{ca}:} The stage task masks the tokens in the target sequence with different tokens based on whether they have ever appeared in the source sequence or not.

    \smallskip\noindent
    \textbf{\Acl{ce}:} The stage task masks the tokens in the target sequence that have already appeared in the source sequence, to drive the model to connect new tokens.
    
    \smallskip\noindent
    \textbf{\Acl{ci}:} The stage task masks the tokens in the target sequence that have never appeared in the source sequence, to drive the model to discard old tokens.
\end{custombox}

\section{Experiments}
\label{sec:experiments}

To check the effectiveness of our approach which takes reusing pretrained language models as a prerequisite, we first experiment with the transfer learning assembly itself and then introduce adaptive schemes, \eg, continuous pretraining and intermediate finetuning.
In the context of the study, we thereby formulate the research questions as follows.

\begin{reqs}
    \item [\req{1}] How well does our transfer learning model perform in code summarization?
    \item [\req{2}] How well does code summarization benefit from the \ac{awgn} refiner?
    \item [\req{3}] How well does code summarization benefit from adaptive continuous pretraining?
    \item [\req{4}] How well does code summarization benefit from adaptive intermediate finetuning?
\end{reqs}

The replication repository is published online \footnote{\url{https://github.com/jianguda/afm4acs}}, including implementations, configuration, results, and the guidance on reproducing experiments.
The implementations are in \python, using \textsc{PyTorch}~\cite{paszke_pytorch_2019} and \textsc{Transformers}~\cite{wolf_transformers_2020}.
The scoring function invokes the third-party evaluation package~\cite{sharma_relevance_2017}.
The experiments are conducted on a machine with an AMD EPYC 7702 CPU with \num{16} \text{GB} RAM, and a single Nvidia Tesla V100 GPU with \num{32} \text{GB} memory.

In the following, we describe the methodology applied to answer the above-mentioned research questions.

\subsection{Baselines}

We compare our model with modern \ac{sota} models, \eg, \basts~\cite{lin_improving_2021} and \sit~\cite{wu_code_2021}.
They both utilize structure information to improve the code summarization and outperformed other prior models but have never been compared to each other.

\basts parses the code into an \ac{ast} and converts it into smaller trees, capturing the local non-linear syntax information from them.
It splits the code snippet based on its blocks in the dominator tree of the \ac{cfg} and generates a split \ac{ast} for each code split.
Each split \ac{ast} will be encoded in the tree-form separately for concatenation with the usual code embedding.

\sit introduces a structure-induced self-attention mechanism to capture the information on syntax structure, data flow as well as data dependency from \ac{ast}.
It parses the code snippet into an \ac{ast}, then transforms the tree into adjacency matrices of three views, and finally combines them evenly to complement the information of sequential code data.

A crucial difference between our transfer learning model and baseline models is that the baseline models are not compatible with adaptive schemes.
It is caused by the extensibility of our model.
We successively use model tags \expttag{\basts} and \expttag{\sit} to indicate the introduced baseline models.

\subsection{Corpus}

In our experiments, we use three datasets in total.
All datasets consist of numerous pairs of code snippets and code comments in \java or \python.
All the experiments are conducted on \java and \python separately.

In code summarization, the commonly used evaluation is the one used in \sit~\cite{wu_code_2021}, in which \java and \python are selected as the representative research targets considering their language popularity and data richness.
We thereby name this benchmark as \sit dataset, and we conduct experiments on that dataset as well.
The other baseline model, \ie, \basts~\cite{lin_improving_2021}, adopted a different benchmark.
For the sake of clarity, we name that benchmark as \basts dataset. Considering the difficulty of reproducing \basts, due to its weak adaptability of the preprocessing pipeline, especially for the performance loss that happens in switching the benchmark, we evaluate our model on the \basts dataset for fair comparisons.

In addition to \basts and \sit datasets, we introduce the CodeSearchNet dataset~\cite{husain_codesearchnet_2019}, shortly referred to as \csn, to support the experiments for adaptive schemes.
The \csn dataset is collected from publicly available open-source repositories in \github.
The full dataset contains over two million snippet-comment pairs spanning multiple programming languages.
To keep consistent with other datasets, we only utilize its \java and \python data.
We follow the given data partition of the \csn dataset but merge the given test set into the given training set, considering we would never use the given test set.

Actually, the \basts dataset originates from earlier data sources~\cite{husain_codesearchnet_2019,hu_deep_2019}, so does the \sit dataset~\cite{hu_deep_2018, wan_improving_2018}.
Even though the \python data in the \basts dataset are taken from the \csn dataset, these two datasets are never used for the same model, so there is no potential data leak.
Meanwhile, all the data in the \sit dataset originates from \github, so there might be a data overlap between \sit and \csn datasets.
However, in our experimental design, we never directly use data pairs in \csn for sequence-to-sequence learning, so no potential data leak exists as well.

As shown in \cref{tab:corpus}, the first column presents datasets, and the second to ninth columns the statistics on the pairs of code snippets and code comments, per language and dataset.

\begin{table}[!tb]
    \caption{Statistics of the snippet-comment pairs in the corpus}
    \label{tab:corpus}
    \centering
    \resizebox{0.8\linewidth}{!}{%
        \sisetup{table-format=6}
\rowcolors{2}{}{gray!10}
\begin{tabular}{
    l SS SS SS SS
}

\hiderowcolors
\toprule

\multirow{2}[2]{*}{\textbf{Dataset}} & \multicolumn{2}{c}{\textbf{\basts}} & \multicolumn{2}{c}{\textbf{\sit}} & \multicolumn{2}{c}{\textbf{\csn}} \\
\cmidrule(lr){2-3} \cmidrule(lr){4-5} \cmidrule(lr){6-7}
& \textbf{\java} & \textbf{\python} & \textbf{\java} & \textbf{\python} & \textbf{\java} & \textbf{\python} \\

\midrule
\showrowcolors

Train & 415395 & 216436 & 69708 & 57203 & 170106 & 265734 \\
Valid & 12885 & 12119 & 8714 & 19067 & 10955 & 14918 \\
Test & 13237 & 12767 & 8714 & 19066 & {--} & {--} \\

\hiderowcolors
\midrule

Total & 441517 & 241322 & 87136 & 95336 & 181061 & 280652 \\

\bottomrule

\end{tabular}

    }
\end{table}

\subsection{Evaluation Metrics}

Considering BLEU~\cite{papineni_bleu_2001}, METEOR~\cite{lavie_meteor_2007}, and ROUGE~\cite{lin_rouge_2004} are the most common evaluation metrics in code summarization, we thus use them to evaluate our experiments~\cite{gros_code_2020}.

The definitions of BLEU, METEOR, and ROUGE are based on the same scenario, \ie, for candidate sentences in a corpus, there are a set of reference sentences that corresponds to each of them.
In their implementations, the penalty factor $\rho$ is for rational adjustments.
Besides, their computations rely on the Precision score and the Recall score, short as $P$ and $R$:

\begin{equation*}
~P_{n}=\frac{\mathsf{\#gram}_{n}(c, r)}{\mathsf{\#gram}_{n}(c)},
~R_{n}=\frac{\mathsf{\#gram}_{n}(c, r)}{\mathsf{\#gram}_{n}(r)}
\end{equation*}

where $c$ and $r$ are the candidate sentence and reference sentence, respectively, and $\mathsf{\#gram}$ is the number of overlapped $n$-grams.

\paragraph{BLEU} The BLEU score computes the averaged percentage of $n$-gram matches between the candidate sentence and the reference sentence, typically unigrams through \num{4}-grams:

\begin{equation}
\text{BLEU} = \rho \cdot \exp \left(\frac{1}{N} \sum_{n=1}^{N} \log P_{n}\right)
\end{equation}

There are two implementation versions of the BLEU score.
In one case, as proposed initially, the computation is on the corpus level, namely computing one score for the whole corpus, the score is \enquote{corpus BLEU}, referred to as C-BLEU.
In the other case, the computation is on the sentence level, \ie, computing one score for each sentence and taking their arithmetic mean as the final score, the score is \enquote{sentence BLEU}, referred to as S-BLEU.

\paragraph{METEOR} The METEOR score builds alignments on unigrams between the candidate sentence and the reference sentence, but during the process, alignments are prioritized on the longer $n$-grams:

\begin{equation}
\text{METEOR} = \rho \cdot \frac{10 {P_{n}}{R_{n}}}{{R_{n}}+{9 P_{n}}}
\end{equation}

Roughly speaking, the penalty factor depends on the actual alignment.
When the alignment is mainly on the unigrams, the penalty could be rather heavy because of the low similarity.

\paragraph{ROUGE} The ROUGE score counts the overlaps of $n$-grams or \ac{lcs} between the candidate sentence and the corresponding reference sentence.
Taking the $n$-grams as an example, the formula is as follows:

\begin{equation}
\text{ROUGE} = \frac{2 {P_{n}}{R_{n}}}{{R_{n}}+{P_{n}}}
\end{equation}

The most common ROUGE scores are ROUGE-\num{1}, ROUGE-\num{2}, and ROUGE-L.
Their distinction lies on the overlapped target to be counted, \eg, unigrams, \num{2}-grams, or \ac{lcs}.

In our experiments, we compute C-BLEU, S-BLEU, METEOR, and ROUGE-L to quantify the quality of the summarization results.
These metrics are in the range of $[0, 1]$ and will be reported in the percentage form.
The larger the value, the better the effect.
Based on the conclusion of the empirical study of automatic evaluation metrics for code summarization, we give priority to METEOR~\cite{roy_reassessing_2021}.

\subsection{Methodology for \req{1}}

There have been numerous pretrained language models proposed for general code-related tasks~\cite{lu_codexglue_2021}, but there is not yet prior work on leveraging existing checkpoints for code summarization.
Meanwhile, what the performance margin is between such transfer learning models and existing well-designed models is still unknown to the community.

First, we design experiments to compare \adamo and \ac{sota} models on the same datasets.
Considering both the \basts and \sit models involve complicated preprocessing steps to extract structure information, it is more fair and intuitive to experiment with \adamo on corresponding datasets directly.
Moreover, the potential performance reduction caused by scripts migration may not be easily avoided.
In detail, we experiment with \adamo on the \basts dataset to compare with \basts, and on the \sit dataset to compare with \sit.

We conducted two experiments to build an intuitive understanding of model performance. In one group of experiments, we directly evaluate the model without the finetuning step.
In this way, we could estimate how well the pretrained model performs in itself, \ie, without any knowledge of the downstream task, namely \emph{zeroshot learning}~\cite{xian_zeroshot_2017}.
In the other group, we evaluate the encoder-decoder model with the finetuning step. Here we assign a roughly equivalent time budget, \eg, \num{24} hours, to \basts, \sit, and \adamo for a fair comparison.

The tag used for zeroshot learning model is \expttag{AdaMo-0shot}, and for the normal finetuning is \expttag{AdaMo-basic}.
In our experiments, the noise emitter is temporarily turned off.

\subsection{Methodology for \req{2}}

To study the effectiveness of the \ac{awgn} emitter and how its configuration affects the performance of our transfer learning model, we enable the Gaussian noise emitter but with different configurations.
Besides, we run each experiment for the same epoch number with the noise emitter off.

The experimental design is almost the same as for \req{1} because we merely adjust the variance to specific values, without making any other changes.
The experiments are on the \sit dataset, which was commonly used in prior work.

The tag used for the noise model is \expttag{AdaMo-noise}.
We use labels in the form of \exptlabel{AdaMo-noise}{$\sigma$} to distinguish different settings, where the standard deviation \expttag{$\sigma$} could be \expttag{0.1}, \expttag{0.2}, \expttag{0.3}, \expttag{0.4} or \expttag{0.5}.
For example, the label for the \ac{awgn} emitter, when the standard deviation $\sigma$ is \expttag{0.1}, is \exptlabel{AdaMo-noise}{0.1}.

\subsection{Methodology for \req{3}}

Results from \ac{nlp} indicate that the second phase of pretraining in the domain leads to performance gains~\cite{gururangan_don_2020}.
Therefore, it has a promising potential improvement for code summarization.

As introduced in \cref{subsec:as}, we adopt \acf{da} and \acf{ta} pretraining as the additional procedure of tailoring the pretrained model to the data of target domains or tasks.
For \ac{da} pretraining, we utilize the \csn dataset to train encoder, decoder, or them both separately.
For \ac{ta} pretraining, we separately train encoder, decoder, or both on the \sit dataset.
The \basts dataset is excluded since it has a times amount of \java data to \csn and their \python data show an extreme overlap.

As the experimental settings of continuous pretraining, we train the model in both the \ac{da} and \ac{ta} way.
Meanwhile, we separately train the encoder, the decoder, and them both in all experiments.
To check the performance of the adaptive model after the adaptive continuous pretraining scheme, we offer \num{24} hours for the additional training phase.
That is, the encoder, the decoder, or them both are pretrained for \num{24} hours first, and then \adamo itself is trained for \num{24} hours.
Meanwhile, to validate whether continuous pretraining usually requires a longer time like the normal pretraining, we offer a richer time budget, \eg, \num{48} hours, as the reference.

The tag used for continuous pretraining is \expttag{AdaMo-CP}. We use labels in the form of \exptlabelo{AdaMo-CP}{ADAPTIVE}{OBJ} to distinguish different settings, where \expttag{ADAPTIVE} can be \expttag{DA} or \expttag{TA}, and \expttag{OBJ} could be \expttag{mlm}, \expttag{clm} or \expttag{both}.
For example, the label of continuous pretraining only the encoder in the domain-adaptive manner is referred to as \exptlabelo{AdaMo-CP}{DA}{mlm}.

\subsection{Methodology for \req{4}}

For \ac{nlp} tasks that require high-level inference and reasoning capabilities, it proves to be effective to introduce relevant tasks with massive data as the intermediate finetuning~\cite{pruksachatkun_intermediatetask_2020}.

As introduced in \cref{subsec:as}, we want to check whether the intermediate finetuning might benefit our transfer learning approach.
Meanwhile, we expect the ideas of domain and task adaptations could bring potential performance improvements.

As the experimental settings of intermediate finetuning, we conduct two sets of experiments where the model is trained in the domain-adaptive or task-adaptive ways.
In each set, there are three experiments in which \ac{ca}, \ac{ce} or \ac{ci} is adopted as the intermediate stage task.
We train the model for the stage task on the \csn dataset as the domain-adaptive setting, and on the \sit dataset as the task-adaptive setting.
The \basts dataset is excluded since it has a times amount of \java data to \csn and their \python data show an extreme overlap.

To investigate the performance of \adamo after the intermediate finetuning scheme, we offer \num{24} hours for the additional training phase.
For the domain-adaptive and task-adaptive settings, they both have additional \num{24} hours for intermediate finetuning, apart from the \num{24} hours for the normal finetuning.

The tag used for intermediate finetuning is \expttag{AdaMo-IF}.
We use labels in the form of \exptlabelo{AdaMo-IF}{ADAPTIVE}{PROXY} to distinguish different settings, where \expttag{ADAPTIVE} could be \expttag{DA} or \expttag{TA}, and \expttag{PROXY} could be \expttag{CA}, \expttag{CE} or \expttag{CI}.
For example, the label of the \ac{ca} stage task for domain-adaptive is referred to as \exptlabelo{AdaMo-IF}{DA}{CA}.

\section{Results}
\label{sec:results}

In this section, we present the results of our experiments to answer the research questions.
In the following comparisons, the most competitive results are highlighted in bold.

\subsection{Results of \req{1}}

The results for \req{1} are the summarization scores of our basic transfer learning approach, in comparison with the baseline models introduced above.

\begin{table}[!tb]
    \caption{Comparative results on code summarization}
    \label{tab:results_baseline}
    \centering
    \resizebox{\linewidth}{!}{%
        \sisetup{table-format=2.2}
\rowcolors{2}{}{gray!10}
\begin{tabular}{
    ll SSSS SSSS
}

\hiderowcolors
\toprule

\multirow{2}[2]{*}{\textbf{Dataset}} & \multirow{2}[2]{*}{\textbf{Model}} & \multicolumn{4}{c}{\textbf{\java}} & \multicolumn{4}{c}{\textbf{\python}} \\
\cmidrule(lr){3-6} \cmidrule(lr){7-10}
& & {\textbf{C-BLEU}} & {\textbf{S-BLEU}} & {\textbf{METEOR}} & {\textbf{ROUGE}} & {\textbf{C-BLEU}} & {\textbf{S-BLEU}} & {\textbf{METEOR}} & {\textbf{ROUGE}} \\

\midrule
\showrowcolors

\basts & \expttag{\basts} & 27.82\% & 34.22\% & 22.86\% & 45.78\% & 02.33\% & 14.18\% & 08.65\% & 20.87\% \\
& \expttag{AdaMo-0shot} & 00.00\% & 01.80\% & 00.10\% & 00.24\% & 00.00\% & 01.78\% & 00.15\% & 00.36\% \\
& \expttag{AdaMo-basic} & \tabhvalue 31.38\% & \tabhvalue 37.64\% & \tabhvalue 25.59\% & \tabhvalue 49.90\% & \tabhvalue 05.19\% & \tabhvalue 16.46\% & \tabhvalue 12.51\% & \tabhvalue 27.31\% \\

\midrule

\sit & \expttag{\sit} & 38.01\% & 44.96\% & 27.09\% & 53.54\% & 26.46\% & 33.81\% & 21.37\% & 41.18\% \\
& \expttag{AdaMo-0shot} & 00.00\% & 01.79\% & 00.12\% & 00.22\% & 00.00\% & 01.89\% & 00.07\% & 00.13\% \\
& \expttag{AdaMo-basic} & \tabhvalue 40.49\% & \tabhvalue 45.30\% & \tabhvalue 28.19\% & \tabhvalue 53.99\% & \tabhvalue 26.52\% & \tabhvalue 33.85\% & \tabhvalue 21.68\% & \tabhvalue 41.25\% \\

\bottomrule

\end{tabular}

    }
\end{table}

As shown in \cref{tab:results_baseline}, on both \basts and \sit datasets, \expttag{AdaMo-0shot} reaches the lowest scores.
On the contrary, \expttag{AdaMo-basic} always performs better than baseline models on all metrics.
On the one hand, it shows the incoordination of parallel representations damages the zeroshot ability of \expttag{AdaMo}.
On the other hand, it proves the strategy of directly reusing the well-behaved model structures and their trained weights is effective, meanwhile, the training work of \expttag{AdaMo-basic} is not merely for tuning the encoder and decoder, but also to overcome the incoordination issue of parallel representations.

Considering that \basts and \sit are separately reproduced in their respective dataset, it is infeasible to compare their performance directly.
With the reference to \expttag{AdaMo-basic}, it outperforms \basts significantly but only obtains moderate advantages over \sit, therefore the performance of \sit is closer to \adamo than \basts, in terms of the model effects.

\begin{custombox}[\req{1} -- Takeaway]
    Our transfer learning approach beats the \ac{sota} models smoothly when given the same time budget.
    Even though \adamo reuses the existing language models and their trained weights, it is indispensable to train their assembly on the data of target domain or for the target task.
\end{custombox}

\subsection{Results of \req{2}}

The results for \req{2} are the summarization scores of \adamo accumulating varied intensities of Gaussian noise, compared with our basic transfer learning approach.

\begin{table}[!tb]
    \caption{Results of the model with different \ac{awgn} refiners}
    \label{tab:results_refiner}
    \centering
    \resizebox{\linewidth}{!}{%
        \sisetup{table-format=2.2}
\rowcolors{2}{}{gray!10}
\begin{tabular}{
    l SSSS SSSS
}

\hiderowcolors
\toprule

\multirow{2}[2]{*}{\textbf{Model}} & \multicolumn{4}{c}{\textbf{\java}} & \multicolumn{4}{c}{\textbf{\python}} \\
\cmidrule(lr){2-5} \cmidrule(lr){6-9}
& {\textbf{C-BLEU}} & {\textbf{S-BLEU}} & {\textbf{METEOR}} & {\textbf{ROUGE}} & {\textbf{C-BLEU}} & {\textbf{S-BLEU}} & {\textbf{METEOR}} & {\textbf{ROUGE}} \\

\midrule
\showrowcolors

\expttag{AdaMo-basic} & 40.49\% & 45.30\% & 28.19\% & 53.99\% & 26.52\% & 33.85\% & 21.68\% & 41.25\% \\
\exptlabel{AdaMo-noise}{0.1} & 39.04\% & 43.41\% & 26.86\% & 51.67\% & 25.21\% & 32.55\% & 20.79\% & 39.45\% \\
\exptlabel{AdaMo-noise}{0.2} & \tabhvalue 40.62\% & 45.33\% & 28.18\% & 53.86\% & 26.65\% & 34.03\% & \tabhvalue 21.95\% & \tabhvalue 41.83\% \\
\exptlabel{AdaMo-noise}{0.3} & 40.52\% & \tabhvalue 45.35\% & \tabhvalue 28.25\% & \tabhvalue 54.06\% & \tabhvalue 26.80\% & \tabhvalue 34.05\% & 21.92\% & 41.67\% \\
\exptlabel{AdaMo-noise}{0.4} & 40.37\% & 44.92\% & 27.93\% & 53.36\% & 26.78\% & 34.04\% & 21.89\% & 41.71\% \\
\exptlabel{AdaMo-noise}{0.5} & 40.56\% & 45.00\% & 28.04\% & 53.49\% & 26.74\% & \tabhvalue 34.05\% & 21.91\% & 41.64\% \\

\bottomrule

\end{tabular}
    }
\end{table}

As shown in \cref{tab:results_refiner}, the effectiveness of the \ac{awgn} emitter relies on whether or not the noise emitter is well configured.
We observed that with increasing values of $\sigma$ until it is \expttag{0.5}, the effects decrease first and then increase.
When the standard deviation $\sigma$ is \expttag{0.3}, \adamo obtains the best performance.
After that, the results start to fluctuate and enter the downtrend on \java or a stable state on \python.

Comparing the effects of Gaussian noise on specific languages, the results of \expttag{AdaMo-basic} improve on all evaluation metrics only when the $\sigma$ is set to \expttag{0.3} on \java.
On the contrary, results of almost all versions of \expttag{AdaMo-noise} are always better than \expttag{AdaMo-basic} on \python.
The cause is believed to lie in the differences in language expressiveness.
The code snippets in \python are more intuitive than those in \java, and also the \python programs are more close to the common English expressions.
English is a natural language while programming languages are artificial or constructed languages.
Considering that \ac{awgn} is commonly used to mimic the random processes in nature and that \python programs are more close to the English text than \java in terms of coding rules, the Gaussian noise should be more effective when the data itself is more natural, or less artificial.

Overall, the improvements brought by the Gaussian noise are reliable, but it is not conclusive enough.
One reason could be the distinctions in data characteristics.
The Gaussian noise could not perfectly simulate the context information of the code data, as it does in the text data.
The code data is more artificial while the text data is more natural, therefore the effectiveness of Gaussian noise is reduced.
The other reason could be the natural difference between the machine translation task and the code summarization task.
Code summarization is a hybrid task of text summarization and machine translation because the code snippets are strictly written in compiler-oriented grammar rules, but the code comments flexibly follow the natural grammar rules, and meanwhile, the information that existed in the source data is largely reduced after summarization.

\begin{custombox}[\req{2} -- Takeaway]
    The Gaussian noise benefits the model when the noise emitter is well configured. The effects are more accessible if the data itself is closer to the natural language.
    However, the improvements of \ac{awgn} are reliable for code summarization but not very obvious.
\end{custombox}

\subsection{Results of \req{3}}

The results for \req{3} show the summarization scores of \adamo when applying the continuous pretraining scheme, by comparing with our basic transfer learning approach.
As we offer two different time budgets, the falling scores for the same environmental setting are marked with an asterisk.

\begin{table}[!tb]
    \caption{Results of the continually pretraining scheme (\num{24} hours)}
    \label{tab:results_pretrain}
    \centering
    \resizebox{\linewidth}{!}{%
        \sisetup{table-format=2.2}
\rowcolors{2}{}{gray!10}
\begin{tabular}{
    l SSSS SSSS
}

\hiderowcolors
\toprule

\multirow{2}[2]{*}{\textbf{Model}} & \multicolumn{4}{c}{\textbf{\java}} & \multicolumn{4}{c}{\textbf{\python}} \\
\cmidrule(lr){2-5} \cmidrule(lr){6-9}
& {\textbf{C-BLEU}} & {\textbf{S-BLEU}} & {\textbf{METEOR}} & {\textbf{ROUGE}} & {\textbf{C-BLEU}} & {\textbf{S-BLEU}} & {\textbf{METEOR}} & {\textbf{ROUGE}} \\

\midrule
\showrowcolors

\expttag{AdaMo-basic} & 40.49\% & \tabhvalue 45.30\% & 28.19\% & \tabhvalue 53.99\% & 26.52\% & 33.85\% & 21.68\% & 41.25\% \\
\exptlabelo{AdaMo-CP}{DA}{mlm} & 40.50\% & 45.05\% & 27.98\% & 53.40\% & 26.47\% & 33.83\% & 21.55\% & 40.99\% \\
\exptlabelo{AdaMo-CP}{DA}{clm} & 40.18\% & 45.22\% & \tabhvalue 28.20\% & 53.94\% & 26.35\% & 33.68\% & 21.73\% & 41.41\% \\
\exptlabelo{AdaMo-CP}{DA}{both} & \tabhvalue 40.62\% & 45.05\% & 28.16\% & 53.82\% & 26.35\% & 33.61\% & 21.46\% & 40.98\% \\
\exptlabelo{AdaMo-CP}{TA}{mlm} & 40.61\% & 45.27\% & 28.08\% & 53.72\% & \tabhvalue 27.10\% & \tabhvalue 34.43\% & \tabhvalue 22.25\% & \tabhvalue 42.53\% \\
\exptlabelo{AdaMo-CP}{TA}{clm} & 40.01\% & 44.97\% & 28.19\% & 53.96\% & 26.53\% & 33.80\% & 21.89\% & 41.66\% \\
\exptlabelo{AdaMo-CP}{TA}{both} & 40.45\% & 44.93\% & 27.98\% & 53.64\% & 26.86\% & 34.21\% & 22.16\% & 42.37\% \\

\bottomrule

\end{tabular}

    }
\end{table}

Based on the experimental results shown in \cref{tab:results_pretrain}, the continual pretraining scheme shows poor performance on \java but could reliably improve the results on \python.

If we compare the results of \exptlabel{AdaMo-CP}{DA} with \exptlabel{AdaMo-CP}{TA}, then task-adaptive models usually perform more satisfying than domain-adaptive ones. Even though the domain-adaptive scheme owns the advantage in resource intensiveness, the task-adaptive one shows higher efficiency in data utilization.
Therefore, we conclude that task-adaptive is more suitable for continual pretraining.

When other experimental settings are the same, pretraining only the decoder seems the optimal choice since the results of \exptlabelo{AdaMo-CP}{DA}{clm} is better than those of other \exptlabel{AdaMo-CP}{DA} models, so does \exptlabelo{AdaMo-CP}{TA}{mlm}.
A special case is \exptlabelo{AdaMo-CP}{TA}{mlm}, which performs better on \python.

The phenomenon can be explained by two reasons.
First, code data follow strict grammar rules than text data; therefore the text data contain more extensive entropy, which indicates the decoder learning the interrelations of text tokens is more efficient.
Second, when it is task-adaptive, the representation of \python code could be easier to be optimized by the encoder.

\begin{table}[!tb]
    \caption{Results of the continually pretraining scheme (\num{48} hours)}
    \label{tab:results_pretrain2}
    \centering
    \resizebox{\linewidth}{!}{%
        \sisetup{table-format=2.2}
\rowcolors{2}{}{gray!10}
\begin{tabular}{
    l SSSS SSSS
}

\hiderowcolors
\toprule

\multirow{2}[2]{*}{\textbf{Model}} & \multicolumn{4}{c}{\textbf{\java}} & \multicolumn{4}{c}{\textbf{\python}} \\
\cmidrule(lr){2-5} \cmidrule(lr){6-9}
& {\textbf{C-BLEU}} & {\textbf{S-BLEU}} & {\textbf{METEOR}} & {\textbf{ROUGE}} & {\textbf{C-BLEU}} & {\textbf{S-BLEU}} & {\textbf{METEOR}} & {\textbf{ROUGE}} \\

\midrule
\showrowcolors

\expttag{AdaMo-basic} & 40.49\% & 45.30\% & 28.19\% & 53.99\% & 26.52\% & 33.85\% & 21.68\% & 41.25\% \\
\exptlabelo{AdaMo-CP}{DA}{mlm} & \tabhvalue 40.61\% & 45.18\% & 28.03\% & 53.49\% & 26.28\%* & 33.56\%* & 21.46\%* & 40.85\%* \\
\exptlabelo{AdaMo-CP}{DA}{clm} & 39.82\%* & 44.97\%* & 28.10\%* & 53.95\% & 26.34\%* & 33.67\%* & 21.81\% & 41.69\% \\
\exptlabelo{AdaMo-CP}{DA}{both} & 40.39\%* & 44.94\%* & 27.97\%* & 53.40\%* & 26.50\% & 33.76\% & 21.78\% & 41.53\% \\
\exptlabelo{AdaMo-CP}{TA}{mlm} & 40.55\%* & 45.37\% & 28.15\% & 53.97\% & 26.96\%* & \tabhvalue 34.39\%* & 22.08\%* & 42.18\%* \\
\exptlabelo{AdaMo-CP}{TA}{clm} & 40.49\% & \tabhvalue 45.46\% & \tabhvalue 28.36\% & \tabhvalue 54.27\% & 26.61\% & 33.86\% & 21.84\%* & 41.58\%* \\
\exptlabelo{AdaMo-CP}{TA}{both} & 40.51\% & 45.09\% & 28.07\% & 53.68\% & \tabhvalue 27.03\% & 34.30\% & \tabhvalue 22.25\% & \tabhvalue 42.49\% \\

\bottomrule

\end{tabular}

    }
\end{table}

As shown in \cref{tab:results_pretrain2}, if we assign a richer time budget for the continuous pretraining scheme, the patterns found before keep constant.
Besides, all best results are contributed by \ac{cp} models.
Meanwhile, we found that a long training time seems not necessary, since almost half of the scores are reduced a bit.
Overall, that continuously pretraining both encoder and decoder seldom promises better results than training only encoder or decoder.
It should be caused by the incoordination of parallel representations and would be solved after sufficient finetuning.

\begin{custombox}[\req{3} -- Takeaway]
    The effects of continuous pretraining exist but vary with code data.
    Its combination with the task-adaptive scheme promises better improvements.
    It is recommended to only continuously pretrain the decoder since pretraining both the encoder and decoder is not necessarily better.
\end{custombox}

\subsection{Results of \req{4}}

The results for \req{4} are the summarization scores of \adamo when applying the intermediate finetuning scheme, in comparison with our basic transfer learning approach.

\begin{table}[!tb]
    \caption{Results of the intermediate finetuning scheme (\num{24} hours)}
    \label{tab:results_finetune}
    \centering
    \resizebox{\linewidth}{!}{%
        \sisetup{table-format=2.2}
\rowcolors{2}{}{gray!10}
\begin{tabular}{
    l SSSS SSSS
}

\hiderowcolors
\toprule

\multirow{2}[2]{*}{\textbf{Model}} & \multicolumn{4}{c}{\textbf{\java}} & \multicolumn{4}{c}{\textbf{\python}} \\
\cmidrule(lr){2-5} \cmidrule(lr){6-9}
& {\textbf{C-BLEU}} & {\textbf{S-BLEU}} & {\textbf{METEOR}} & {\textbf{ROUGE}} & {\textbf{C-BLEU}} & {\textbf{S-BLEU}} & {\textbf{METEOR}} & {\textbf{ROUGE}} \\

\midrule
\showrowcolors

\expttag{AdaMo-basic} & 40.49\% & 45.30\% & 28.19\% & 53.99\% & 26.52\% & 33.85\% & 21.68\% & 41.25\% \\
\exptlabelo{AdaMo-IF}{DA}{CA} & 29.06\% & 35.75\% & 20.53\% & 44.20\% & 25.60\% & 33.05\% & 21.02\% & 40.11\% \\
\exptlabelo{AdaMo-IF}{DA}{CE} & \tabhvalue 41.21\% & \tabhvalue 46.33\% & \tabhvalue 29.11\% & \tabhvalue 55.59\% & \tabhvalue 28.37\% & \tabhvalue 35.44\% & \tabhvalue 23.25\% & \tabhvalue 44.28\% \\
\exptlabelo{AdaMo-IF}{DA}{CI} & 40.65\% & 45.99\% & 28.83\% & 55.31\% & 27.66\% & 34.88\% & 22.96\% & 43.82\% \\
\exptlabelo{AdaMo-IF}{TA}{CA} & 39.79\% & 44.63\% & 27.73\% & 53.19\% & 25.84\% & 33.42\% & 21.42\% & 40.97\% \\
\exptlabelo{AdaMo-IF}{TA}{CE} & 40.50\% & 45.42\% & 28.02\% & 53.45\% & 26.53\% & 33.91\% & 21.15\% & 40.51\% \\
\exptlabelo{AdaMo-IF}{TA}{CI} & 40.15\% & 45.33\% & 28.30\% & 54.24\% & 26.22\% & 33.65\% & 21.79\% & 41.44\% \\

\bottomrule

\end{tabular}

    }
\end{table}

According to the experimental results shown in \cref{tab:results_finetune}, the combination of domain-adaptive finetuning with concept extrapolation achieves the best scores on all metrics.
In contrast, the combination of domain-adaptive with concept annotation performs the worst and even worse than \expttag{AdaMo-basic}.
All the other combinations have similar performance to \expttag{AdaMo-basic}, with slight improvements or a bit deteriorations.

When comparing domain-adaptive with task-adaptive, the former performs better than the latter if the intermediate stage task is \ac{ce} or \ac{ci}, but the situation becomes opposite if the stage task is \ac{ca}.
We believe that \ac{ce} and \ac{ci} are complex tasks while \ac{ca} is rather a simple task.
The domain-adaptive scheme learns from more data compared to the task-adaptive one, therefore it shows better generalization when given the same time budget.
However, simple intermediate tasks might mislead the domain-adaptive scheme.
The opposite case is that the performance of the task-adaptive scheme almost keeps unchanged no matter which stage task it is.
Therefore, the domain-adaptive scheme should pair with complex intermediate tasks, while task-adaptive might not benefit too much from the intermediate finetuning.

\begin{table}[!tb]
    \caption{Showcase of the effects of the intermediate tasks}
    \label{tab:showcase}
    \centering
    \resizebox{\linewidth}{!}{%
        \sisetup{table-format=2.2}
\rowcolors{2}{}{gray!10}
\begin{tabular}{
    lll
}

\hiderowcolors
\toprule

\textbf{Sample} & \textbf{Model} & \textbf{Comment} \\

\midrule
\showrowcolors

\expttag{Java\#229} & \expttag{AdaMo-basic} & {returns true if the given word contains a whitespace} \\
& \exptlabelo{AdaMo-IF}{DA}{CE} & {returns true \emph{if the input string contains a word} except for word engines} \\
& \exptlabelo{AdaMo-IF}{DA}{CI} & {check \emph{if the input string contains a word}} \\
& \textbf{Ground Truth} & {returns true \emph{if the input string contains a word} - breaking character} \\

\midrule

\expttag{Java\#517} & \expttag{AdaMo-basic} & {ensures that the object value is at the given allocation} \\
& \exptlabelo{AdaMo-IF}{DA}{CE} & {ensures that a value \emph{is not null}} \\
& \exptlabelo{AdaMo-IF}{DA}{CI} & {ensures that the given location \emph{is not null}} \\
& \textbf{Ground Truth} & {ensure the given value \emph{is not null} and return it} \\

\midrule

\expttag{Java\#821} & \expttag{AdaMo-basic} & {draws a face of the specified shape} \\
& \exptlabelo{AdaMo-IF}{DA}{CE} & {draws a circle for \emph{the given parameters}} \\
& \exptlabelo{AdaMo-IF}{DA}{CI} & {draws a dial on \emph{the given parameters}} \\
& \textbf{Ground Truth} & {draws a cylinder for \emph{the given parameters}} \\

\midrule

\expttag{Python\#201} & \expttag{AdaMo-basic} & {remove the most recent history from the history} \\
& \exptlabelo{AdaMo-IF}{DA}{CE} & {remove all \emph{completed jobs}} \\
& \exptlabelo{AdaMo-IF}{DA}{CI} & {remove all the \emph{completed jobs} from the history} \\
& \textbf{Ground Truth} & {remove all \emph{completed jobs} from history} \\

\midrule

\expttag{Python\#439} & \expttag{AdaMo-basic} & {return a string description of the appropriate description on the given path} \\
& \exptlabelo{AdaMo-IF}{DA}{CE} & {\emph{return a string describing the probable encoding of a} file} \\
& \exptlabelo{AdaMo-IF}{DA}{CI} & {\emph{return a string describing the probable encoding of a} unicode path} \\
& \textbf{Ground Truth} & {\emph{return a string describing the probable encoding of a} file} \\

\midrule

\expttag{Python\#660} & \expttag{AdaMo-basic} & {returns a set of all cliques of a chordal graph} \\
& \exptlabelo{AdaMo-IF}{DA}{CE} & {return the \emph{set of maximal cliques of} the chordal graph} \\
& \exptlabelo{AdaMo-IF}{DA}{CI} & {returns \emph{set of maximal cliques of} the chordal graph g} \\
& \textbf{Ground Truth} & {returns the \emph{set of maximal cliques of} a chordal graph} \\

\bottomrule

\end{tabular}

    }
\end{table}

Among all intermediate tasks, concept annotation is weaker than others.
Meanwhile, \ac{ca} is sensitive to the data in some cases, such as pairing with the domain-adaptive scheme on the \java data.
However, other tasks usually promise better results and are insensitive to languages.
On both \java and \python, \ac{ce} and \ac{ci} obtain at least similar but usually better results than \expttag{AdaMo-basic}, therefore, complex intermediate tasks are more stable and beneficial than simple ones.
When pairing with the domain-adaptive scheme, concept extrapolation always performs better than concept interpolation, and while pairing the task-adaptive scheme, their results are close.

To check the effects of intermediate tasks on the generated comments, especially \ac{ce} and \ac{ci}, we could observe selected examples as shown in \cref{tab:showcase}. The longest common substrings occurred in others that include the ground truth, but not in \expttag{AdaMo-basic}, are emphasized in italics, and they contribute the most to the improvements.
Besides, the distinctions between \ac{ce} and \ac{ci} cause differences in other tokens, which are usually never appeared in the corresponding code data for the former \ac{ce} case but often already appeared there for the latter \ac{ci} case.

\begin{custombox}[\req{4} -- Takeaway]
    The combination of complex intermediate tasks with the domain-adaptive scheme is the best choice.
    The optimal intermediate task is concept extrapolation.
    The task-adaptive scheme promises more stable results, but its improvements are not so obvious as for the domain-adaptive scheme.
\end{custombox}

\subsection{Discussion}

To build a clear comparison of the effect promotion given by each extension, we compute the growth extents of best scores in each research question relative to those of \sit, as shown in \cref{tab:promotion}.
Summarizing the results, we notice that the intermediate finetuning scheme pairing with domain-adaptive always boosts the results.
On \java, the second largest growth origins from its combination with the task-adaptive way.
On \python, the second-best results are contributed by the continual pretraining scheme pairing with the task-adaptive way.
Then, the refiner with adequate Gaussian noise performs the optimal while others have close results with the basic \adamo.

\begin{table}[!tb]
    \caption{Overview of the promotion effects on code summarization}
    \label{tab:promotion}
    \centering
    \resizebox{\linewidth}{!}{%
        \sisetup{table-format=2.2}
\rowcolors{2}{}{gray!10}
\begin{tabular}{
    l SSSS SSSS
}

\hiderowcolors
\toprule

\multirow{2}[2]{*}{\textbf{Model}} & \multicolumn{4}{c}{\textbf{\java}} & \multicolumn{4}{c}{\textbf{\python}} \\
\cmidrule(lr){2-5} \cmidrule(lr){6-9}
& {\textbf{C-BLEU}} & {\textbf{S-BLEU}} & {\textbf{METEOR}} & {\textbf{ROUGE}} & {\textbf{C-BLEU}} & {\textbf{S-BLEU}} & {\textbf{METEOR}} & {\textbf{ROUGE}} \\

\midrule
\showrowcolors

\expttag{AdaMo-basic} & 06.52\% & 00.76\% & 04.06\% & 00.84\% & 00.23\% & 00.12\% & 01.45\% & 00.17\% \\
\expttag{AdaMo-noise} & 06.87\% & 00.87\% & 04.28\% & 00.97\% & 01.28\% & 00.71\% & 02.71\% & 01.58\% \\
\exptlabel{AdaMo-CP}{DA} & 06.87\% & 00.58\% & 04.10\% & 00.75\% & 00.04\% & 00.06\% & 01.68\% & 00.56\% \\
\exptlabel{AdaMo-CP}{TA} & 06.84\% & 00.69\% & 04.06\% & 00.78\% & 02.42\% & 01.83\% & 04.12\% & 03.28\% \\
\exptlabel{AdaMo-IF}{DA} & \tabhvalue 08.42\% & \tabhvalue 03.05\% & \tabhvalue 07.46\% & \tabhvalue 03.83\% & \tabhvalue 07.22\% & \tabhvalue 04.82\% & \tabhvalue 08.80\% & \tabhvalue 07.53\% \\
\exptlabel{AdaMo-IF}{TA} & 06.55\% & 01.02\% & 04.47\% & 01.33\% & 00.26\% & 00.30\% & 01.97\% & 00.63\% \\

\bottomrule

\end{tabular}

    }
\end{table}

It seems hard to understand that, when it is domain-adaptive, intermediate finetuning performs better than continuous pretraining.
It seems that \ac{cp} learns the latent representation by optimizing the self-supervised objectives, but \ac{if} learns the interrelations of tokens via our stage tasks in a supervised way.
Therefore, \ac{cp} is slower than \ac{if} when given the same time budget.
When it is task-adaptive, \ac{cp} and \ac{if} show their advantages on separate datasets.
We believe it is due to the fact that the knowledge learned from the target dataset is more easily beneficial to the target task.
Thereby, the results are likely to be intervened by data characteristics and fortuity.

As reflected in our results, Gaussian noise and continuous pretraining have a small contribution to the achieved effectiveness.
There are no alternatives found yet to the \ac{awgn} emitter.
We have experimented with other common noises in signal processing but only got negative results.
For continuous pretraining, our results show that training the encoder did not help a lot, which indicates the \ac{mlm} objective could be replaced with other token-level ones~\cite{aroca-ouellette_losses_2020} for potential improvements.

\subsection{Threats to Validity}

\paragraph{Internal Validity}
The most crucial limitation to our results comes from the representativeness of our evaluation data, although we already use the most common dataset for evaluation.
In our early-stage experiments, we found that results for both baseline models and our transfer learning model, evaluated on the \csn data, are rather low, even though ours are still better.
Based on our analysis to the phenomenon, we concluded that it is caused by the quality differences of code comments.
Therefore, it would be meaningful to evaluate the performance of models on the data of various quality levels systemically.

\paragraph{External Validity}
The results are limited in the way that they can be generalized to relevant generative tasks.
Our approach applies to generative tasks where the input data and output data are sequential data, such as program migration and code generation, respectively from code to code and from text to code.
The limitation is whether or not there are already pretrained models available for generating the data.
However, the task that involves generating code is harder because the generated programs might be partial, buggy, and even specious.
Therefore, there still exists potential challenges on how to effectively generalize our approach.

\section{Related Work}
\label{sec:related_work}

In the following, we discuss prior works on code summarization and summarize the research status in topics of text summarization and program representation.

\paragraph{Code Summarization}
In recent years, there have been several neural-based approaches proposed for code summarization.
\textsc{CODE-NN}~\cite{iyer_summarizing_2016} uses the \ac{lstm} network~\cite{hochreiter_long_1997} combined with global attention~\cite{luong_effective_2015} for both code retrieval and code summarization.
\textsc{Hybrid-DRL}~\cite{wan_improving_2018} applies reinforcement learning to incorporate the \ac{ast} structure and sequential content of snippets by using an actor-critic network.

Besides the token sequence, the parsing results of the code data, \eg, \ac{ast} and \ac{cfg}, are used for code summarization as well.
\textsc{Hybrid-DeepCom}~\cite{hu_deep_2019} fuses the lexical and syntactical information of code tokens and serialized \acp{ast} using the \ac{gru} network~\cite{cho_learning_2014}.
\textsc{AttendGru}~\cite{leclair_neural_2019} applies the \ac{gru} encoders for code sequences and the serialized \acp{ast}, and adopts attention components for their interrelations with the summary tokens.

Not only do neural models, but also information retrieval methods perform well in code summarization, therefore some work is inspired to combine them.
\textsc{Rencos}~\cite{zhang_retrievalbased_2020} utilizes the search engine to find the most semantically or syntactically similar code snippets to augment samples.
\textsc{Hybrid GNN}~\cite{liu_retrievalaugmented_2021} constructs code property graph and meanwhile learns attention-based dynamic graph as the training data.
In the process, it retrieves the most similar code and corresponding summary for information augmentation.

The latest models are focused on adapting Transformer, considering its success in natural language-related tasks.
\textsc{C2NL}~\cite{ahmad_transformerbased_2020} is the enhanced \transformer~\cite{vaswani_attention_2017} equipped with copy attention~\cite{see_get_2017} and relative position encoding~\cite{shaw_selfattention_2018}.
It merely relies on the knowledge from the sequence of code tokens.
Furthermore, \sit~\cite{wu_code_2021} introduces the structure-induced attention mechanism to capture information from syntax structure, data flow, and data dependency.
Follow the idea of building representations for AST~\cite{zhang_novel_2019}, \basts~\cite{lin_improving_2021} decomposes the code data into blocks in \ac{cfg} to generate a split \ac{ast} for each of them and eventually generate corresponding representations.
Therefore, it utilizes the syntax information from split \acp{ast}, instead of the only original \ac{ast}.
Apart from these models, some work systematically studies current issues in code summarization, including the commonly used datasets and evaluation metrics~\cite{gros_code_2020,roy_reassessing_2021}.

\paragraph{Text Summarization}
Inspired by neural machine translation~\cite{sutskever_sequence_2014,bahdanau_neural_2015}, the sequence-to-sequence model with attention is proposed for abstractive summarization~\cite{rush_neural_2015}.
By leveraging the pointer network~\cite{vinyals_pointer_2015}, \textsc{PGNet}~\cite{see_get_2017} drives a pointer and a generator in parallel to freely choose from either the copied contents or the generated tokens.
Meanwhile, it introduces the coverage mechanism~\cite{tu_modeling_2016} to penalize repetitions.
Even further, \textsc{Bottom-Up}~\cite{gehrmann_bottomup_2018} first selects potential tokens for the summary and then generates the summary using the \textsc{PGNet}.
\textsc{SeqCopyNet}~\cite{zhou_sequential_2018} extends the copy mechanism, which not merely learns to copy the isolated tokens, but also the subsequences.
\textsc{SAGCopy}~\cite{xu_selfattention_2020} enhances the copy mechanism based on the token importance.
It builds a directed graph and adopts the degree centrality to identify the key tokens.

Gradually, neural models equipped with copy mechanism are replaced by pretrained models, such as \textsc{PEGASUS}~\cite{zhang_pegasus_2020} for abstractive summarization, as well as \textsc{MASS}~\cite{song_mass_2019} and \textsc{BART}~\cite{lewis_bart_2020} for the general sequence-to-sequence tasks.
Based on \transformer and transfer learning, universal models represented by \textsc{T5}~\cite{raffel_exploring_2020} are proposed, which are intended to solve most common \ac{nlp} tasks at once.
As the reflection of text summarization, \textsc{SummEval}~\cite{fabbri_summeval_2021} intends to resolve critical shortcomings in evaluation methods.

\paragraph{Program Representation}
By simply regarding code data as token sequences, self-supervised representation learning could be tailored for code data, such as \codebert~\cite{feng_codebert_2020}, \textsc{Codex}~\cite{chen_evaluating_2021} and \textsc{PLBART}~\cite{ahmad_unified_2021}.
With hypothesis that \emph{\enquote{programs with the same functionality should have similar underlying representations}}, \textsc{ContraCode}~\cite{jain_contrastive_2020} builds representations of program functionalities by learning from contrastive samples~\cite{hadsell_dimensionality_2006}.

Also, it is common to first parse the code data into tree or graph structures for richer information.
There have been many models proposed to learn the parsed results of code data.
\textsc{ASTNN}~\cite{zhang_novel_2019} splits each \ac{ast} into a sequence of small trees for better representations.
\textsc{MRNCS}~\cite{gu_multimodal_2021} recaps serialization schemes on tree structures and categorized them into sampling-based~\cite{alon_general_2018} and traversal-based ones~\cite{hu_deep_2018}.
\textsc{TDLS}~\cite{allamanis_learning_2018} utilizes \textsc{GGNN}~\cite{li_gated_2016} to learn both syntactic and semantic information. 
Opposite to static analysis, \textsc{DyPro}~\cite{wang_learning_2019a} and \textsc{LiGer}~\cite{wang_learning_2019} learn program representations through dynamic executions as well, from the mixture of symbolic and concrete execution traces.

\section{Conclusions}
\label{sec:conclusions}

In this paper, we demonstrated the transfer learning model performs well in code summarization by assembling available foundation models, \codebert and \gpttwo.
Then we utilized Gaussian noise to optimize the latent representation of the assembly by simulating the context-aware settings.
Last, we introduced continuous pretraining and intermediate finetuning as adaptive schemes for optional improvements.
In addition, we proposed concept interpolation and concept extrapolation as the intermediate stage tasks for code summarization and validated their effectiveness. These tasks apply to general sequence-to-sequence learning as well.
Moreover, we experimented with adaptation ideas to tailor foundation models with the data of related domains or designated tasks.

The goal of our work is to show that the transfer learning model based on existing foundation models is rather competitive and could even outperform \ac{sota} models. Moreover, our results showed that neural models regarding code data merely as sequential data could still be powerful enough.
Also, our model is compatible with various adaptive schemes, which promises further improvements in model performance. Compared with \ac{sota} models, \adamo is more friendly to potential rises in quality from either the model side or the data side.

Despite our results and findings, there are still questions waiting to be solved.
For example, it is challenging to align the latent representations of pretrained encoder and decoder models efficiently, or else, it might be possible to make use of pretrained models in a more flexible way, just like playing building blocks.
Besides, it should be valuable to implement the refiner with certain neural models for the manipulation of latent representations.

\clearpage
\bibliography{references, IEEEsettings}

\end{document}